\documentclass[12pt]{article}
\usepackage{amssymb,dsfont,amsthm,amsmath,graphics,colortbl,epsfig,eepic}

\def\openone{\leavevmode\hbox{\small1\kern-3.3pt\normalsize1}}

\def\bu{{\boldsymbol u}}

\def\tr{\mathrm{tr\,}}

\def\ad{\mathrm{ad\,}}

\def\diag{\mbox{diag\,}}

\def\bbbp{{\Bbb P}}

\def\bbbc{{\Bbb C}}
\def\bbbr{{\Bbb R}}
\def\bbbz{{\Bbb Z}}

\def\rmi{\mathrm{i}}
\def\rmd{\mathrm{d}}

\begin{document}
\arraycolsep=2pt

\numberwithin{equation}{section}
\allowdisplaybreaks
\bibliographystyle{plain}

\begin{center}
{\large \bf Rational Bundles and Recursion Operators for Integrable Equations
on {\bf A.III}-type Symmetric Spaces}

\bigskip

{\bf V. S. Gerdjikov$^1$, G. G. Grahovski$^{1,2}$, A. V. Mikhailov$^3$, \\
T. I. Valchev$^1$}

\medskip

{\it $^1$Institute of Nuclear Research and Nuclear Energy,  Bulgarian Academy of Sciences, 72 Tsarigradsko chausee, Sofia 1784, Bulgaria } \\
{\it $^{2}$School of Mathematical Sciences, Dublin Institute of Technology, \\
Kevin Street, Dublin 8, Ireland} \\
{\it $^3$Applied Math. Department, University of Leeds,\\
Woodhouse Lane, Leeds, LS2 9JT, UK}

\medskip

{\small E-mails: gerjikov@inrne.bas.bg, grah@inrne.bas.bg}\\ {\small a.v.mikhailov@leeds.ac.uk, valtchev@inrne.bas.bg}

\end{center}

\begin{abstract}
\noindent We analyze and compare the methods of construction of
the recursion operators for a special class of integrable
nonlinear differential equations related to {\bf A.III}-type
symmetric spaces in Cartan's classification and having additional
reductions.

\end{abstract}

\medskip

\section{Introduction}

Recursion operators $\Lambda_\pm$ play an important role in the theory of the nonlinear evolution equations (NLEE), integrable by the inverse scattering  method. They have been constructed and analyzed for a wide class of Lax operators $L$ and appeared to generate not only the
Lax representations, but also the hierarchy of NLEE's related to a given Lax operator $L$, their conservation laws and their hierarchy of Hamiltonian
structures, see \cite{DrSok*85,brown-bible,blue-bible,GVY*08} and the numerous references therein.
Such operators can be viewed also as the Lax $L$ operator, taken
in the adjoint representation of the relevant Lie algebra $\mathfrak{g}$.

The construction of $\Lambda_\pm$ for Lax operators, whose explicit dependence on the spectral parameter $\lambda$ is
comparatively simple  (say, linear, or quadratic) has been done a long time ago \cite{AKNS*74,KN,GeKh,bjp}.
Furthermore, the completeness property for the set of eigenfunctions of $\Lambda_\pm$ (the `squared solutions' of $L$) is of paramount importance. The completeness of the `squared solutions' plays a  fundamental role
in proving that the ISM is, in fact, a nonlinear analogue of the Fourier transform, which allows one to linearize the NLEE.
Using these relations one is able to derive all fundamental properties of the NLEE on a common basis.

An important tool of extracting new NLEE from a known multi-component ones, is the the reduction group, introduced in
\cite{mik_toda,miktetr,mik}. It led to the discovery of the 2-dimensional Toda field theories \cite{mik,mop1}. Latest developments
of the method led to the discovery of new automorphic Lie algebras and their classification \cite{lomsan,miklom1,miklom}.

The problem of deriving  recursion operators becomes more
difficult when we impose additional reductions on $L$. If this
additional reduction is compatible with $L$, being linear or
quadratic in $\lambda$, the construction of $\Lambda$ is not a
difficult task,  see \cite{GVY*08}. An alternative construction of
$\Lambda$ as formal recursion operator is given in
\cite{sok,golsok}, see also the review papers \cite{ASY*00,MS*09}.
The  effect of the $\bbbz_n$-reduction is as follows: i) the
relevant `squared solutions' have analyticity properties in
sectors of the complex $\lambda$-plane closing angles $\pi/n$; ii)
the grading of the Lie algebra $\mathfrak{g} \equiv
\oplus_{k=0}^{n-1}\mathfrak{g}^{(k)}$ is more involved and as a
consequence the recursion operator is factorized into a product of
$n$ factors $\Lambda =\prod_{k=0}^{n-1}\Lambda_k$, and each of the
factors $\Lambda_k$ maps $\Lambda_k \colon \mathfrak{g}^{(k-1)}
\to \mathfrak{g}^{(k)}$.

The situation becomes more complicated if the additional reduction drastically changes the $\lambda$-dependence of the Lax operator.
Here we address this problem for one of the simplest nontrivial cases when the Lax operator $L$, due to an additional $\bbbz_2$-reduction
changes its $\lambda$-dependence from polynomial into rational one.  The linear in $\lambda$ Lax operator
\begin{equation}\label{eq:Lax0}
L_0\psi \equiv i \partial_x\psi + \lambda L_1 \psi(x,\lambda) = 0,
\end{equation}
was shown to give rise \cite{sok,ours} to the integrable system
\begin{equation}
 \label{eqiso_Int}
i \bu_t=((1-\bu\bu^{\dag})\bu_x)_x\, , \qquad \bu^{\dag} \bu=\openone  ,
\end{equation}
where $\bu$ is $(N-k)\times k$ complex matrix and $\openone $ is the unit matrix. The system (\ref{eqiso_Int}) is $S(U(N-k)\times U(k))$ invariant,
and, in this sense, is  isotropic. In particular, if $k=1$, equation (\ref{eqiso_Int}) can be seen
as a $U(N-1)$ invariant integrable system on $\bbbc\bbbp^{N-1}$.  The corresponding recursion operators, relevant for the fundamental
properties of eq. (\ref{eqiso_Int}), were derived in \cite{JGSP}.

In \cite{ours} we derived new NLEE with additional $\bbbz_2$-reduction, which maps $\lambda \to\lambda^{-1}$. This reduction
is among the simplest nontrivial ones, see \cite{mik_toda,miktetr,mik,miklom1,miklom}. Due to it,
the relevant Lax operator acquires a rational dependence of $\lambda$:
\begin{equation}\label{eq:Lax}
L\psi \equiv i \partial_x\psi + \left(\lambda L_1 + \frac{1}{\lambda} L_{-1}\right)\psi = 0,
\end{equation}
and the simplest NLEE take the form
\begin{equation}\label{nee}
\begin{split}
i u_t&= u_{xx}-(u(u^*u_x+v^*v_x))_x+8 vv^*u,\\
i v_t&=v_{xx}-(v(u^*u_x+v^*v_x))_x-8uu^*v\, ,
\end{split}
\end{equation}
where $u$ and $v$ are functions of $x$ and $t$ subject to the condition:
\begin{equation}\label{eq:uv0}
|u^2| + |v^2| = 1
\end{equation}
i.e. the vector with components $u$ and $v$ sweeps a $3$-dimensional
sphere in $\bbbr^4$. System (\ref{nee})  can also be seen as an anisotropic
deformation of (\ref{eqiso_Int}) with $k=1,N=3$.

Our main aim in the present paper is to present  two ways of
deriving the recursion operators related to the Lax operators with
rational dependence on $\lambda$.  It is a natural extensions of
our previous results \cite{ours,JGSP}. In the next Section 2 we
give preliminaries concerning the spectral properties of $L$
(\ref{eq:Lax}). In  Section 3 we derive the recursion operator
using the G\"urses-Karasu-Sokolov (GKS) method. In Section  4 we
use the Wronskian relations to determine the `squared solutions of
$L$. Next using the gauge covariant approach \cite{AKNS*74,GVY*08}
we introduce the recursion operator as the one for which the
`squared solutions' are eigenfunctions. In the last Section 5 we
give some conclusions, while the Appendix contains some details
from our calculations.

\section{Preliminaries}

In this section we first formulate the basic results from our
previous paper \cite{ours}. There we have shown that eq.
(\ref{nee}), which is naturally related to the  symmetric space
\cite{Hel} of {\bf A.III}-type $SU(3)/S(U(1)\times U(2))$, allows
a Lax representation and can be solved by the inverse scattering
method. The Lax operator $L$ is given by eq. (\ref{eq:Lax}) and
the equation itself is the compatibility condition between $L$ and
the linear operator
\begin{eqnarray}
A\psi&:=&i \partial_t\psi +\left(A_0+\lambda A_1 +
\frac{1}{\lambda} A_{-1} +\lambda^2 A_2 +
\frac{1}{\lambda^2} A_{-2}\right)\psi = \psi f(\lambda),\label{lax_2}
\end{eqnarray}
It is well know  the above mentioned symmetric space is constructed by using
Cartan's involutive automorphism \cite{Hel}, which  induces a
$\bbbz_2$ grading in the underlying Lie algebra
\begin{equation}\label{eq:g01}
\begin{aligned}
\mathfrak{g} &= \mathfrak{g} ^{\bf (0)}\oplus\mathfrak{g}^{\bf (1)}, \qquad J_1=\diag (1,-1,-1) \\
 \mathfrak{g} ^{(0)} &= \{ Y\in \mathfrak{g} \colon J_1 Y J_1 =  Y\}, \qquad
 \mathfrak{g} ^{(1)} = \{ X\in \mathfrak{g} \colon J_1 X J_1 =  -X\}.
\end{aligned}
\end{equation}
Our Lax representation is such that  $ L_{\pm 1}, A_{\pm 1} \in\mathfrak{g}^1$ and $A_0, A_{\pm 2}\in \mathfrak{g}^0$,
for more details see the Appendix. In particular, for $L_1$ and $A_1$ we write
\begin{equation}\label{eq:L1A1}
L_1= \left(\begin{array}{ccc} 0 & u & v \\ u^* & 0 & 0 \\ v^* & 0 & 0
\end{array}\right), \qquad
A_1 = \left(\begin{array}{ccc} 0 & a & b \\ a^* & 0 & 0 \\ b^* & 0 & 0
\end{array}\right).
\end{equation}
which explicitly involves the first two reduction conditions:
\begin{equation}\label{eq:RC1}
L_{\pm 1}^\dag = L_{\pm 1} , \qquad A_{\pm 1}^\dag = A_{\pm 1} , \qquad A_{\pm 2}^\dag = A_{\pm 2}.
\end{equation}
We impose also a third reduction of the form
\begin{equation}
J_2 L_1 J_2 =L _{-1} ,\qquad J_2 A_1 J_2=A_{-1}  ,\qquad J_2 A_2 J_2=A_{-2}
\label{red3}\end{equation}
where $J_2=\diag(1,-1,1)$. At this point, we impose one additional requirement
\[|u|^2+|v|^2=1,\]
that is the vector $(u,v)$ lives in a 3-dimensional sphere in $\bbbr^4$.

The spectral theory of the Lax operator $L$  (\ref{eq:Lax}) substantially depends on the boundary conditions.
We have two natural ways to fix the boundary configurations, namely:
\begin{enumerate}
\item $\lim_{x\to \pm\infty} u(x,t) =1,
\quad \lim_{x\to \pm\infty} v(x,t) =0$
\item $\lim_{x\to \pm\infty} u(x,t) =0,
\quad \lim_{x\to \pm\infty} v(x,t) =1$
\end{enumerate}

\begin{figure}
\centerline{\includegraphics[width=10cm]{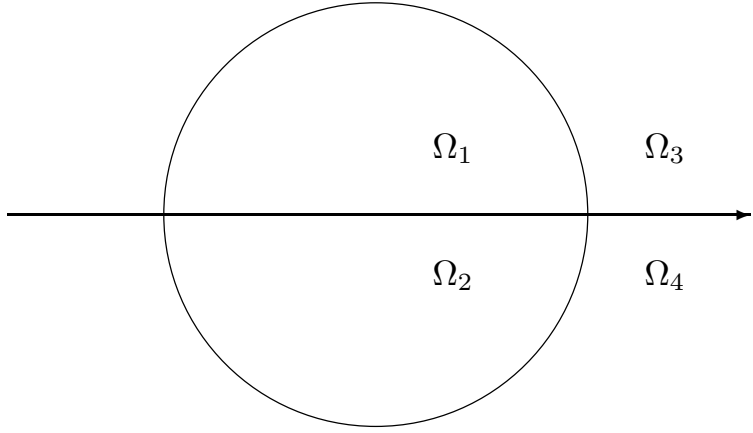}}
\caption{Continuous spectrum of $L$, case b).}\label{spectrum}
\end{figure}

The Jost solutions of $L$ are determined by using the asymptotic potentials
 $U_{\pm, as}(\lambda) = \lim_{x\to\pm\infty} (\lambda L_1 + \lambda^{-1} L_{-1})$ which can be
 diagonalised  for both choices of boundary conditions
\begin{equation}\label{eq:1}
\begin{aligned}
& \mbox{1)} &\qquad g^{-1}_{0}U_{\pm, \rm as}(\lambda) g_{0} &=  J_{\rm a}(\lambda), \qquad
g_{0} = \frac{1}{\sqrt{2}}\left(\begin{array}{ccc} 1 & 0 & 1 \\ 1
& 0 & -1 \\ 0 & -\sqrt{2} & 0 \end{array}\right) \\
& &\qquad J_{\rm a}(\lambda) &= (\lambda -\lambda^{-1}) \diag (1,0,-1)\\
& \mbox{2)} &\qquad g^{-1}_{0}U_{\pm, \rm as} (\lambda)g_{0} &= J_{\rm b}(\lambda),
\qquad g_{0} = \frac{1}{\sqrt{2}}\left(\begin{array}{ccc} 1 & 0 & -1 \\ 0
& \sqrt{2} & 0 \\ 1 & 0 & 1 \end{array}\right) \\
& & \qquad J_{\rm b}(\lambda) &= (\lambda +\lambda^{-1}) \diag (1,0,-1).
\end{aligned}
\end{equation}
The continuous part of the spectrum of $L$ for the case a) fills in the real axis $\bbbr$ while
for the case b) it is the union of the real axis and the unit circle, see Fig.\ref{spectrum}.

In \cite{ours} we have constructed the fundamental analytic solutions (FAS) of $L$ for both cases.
Skipping the details, we note only that these FAS can be viewed as solutions of  Riemann-Hilbert problems (RHP):

The Jost solutions of $L$ and the related scattering matrix $T(\lambda)$ are defined  as follows:
\begin{equation}\label{eq:}
\begin{aligned}
\lim_{x\to\pm\infty}\psi_{\pm}(x,\lambda) e^{-i J(\lambda)x}g_{0}^{-1} &=\openone, \\
T(\lambda) &= \psi_{+}^{-1}(x,\lambda)\psi_{-}(x,\lambda)
\end{aligned}
\end{equation}
For the case a) the fundamental analytic solutions $\chi^{+}$ and $\chi^{-}$ are
solutions of the following RHP
\begin{equation}
\chi^{+}(x,\lambda)=\chi^{-}(x,\lambda)G(\lambda), \quad G(\lambda) =
[S^-(\lambda)]^{-1}S^+(\lambda), \quad \lambda\in \bbbr.
\end{equation}
They obey the symmetry relations imposed by the reductions
\begin{equation}
\begin{split}
(\chi^{+})^\dag (x,\lambda^*) &= \chi^{-}(x,\lambda)\\
J_1\chi^{+}(x,-\lambda)J_1 &= \chi^{-}(x,\lambda)\\
J_2\chi^{\pm}(x,1/\lambda)J_2 &= \chi^{\pm}(x,\lambda).
\end{split}
\end{equation}
In the case b) we have 4 fundamental analytic solutions
$\chi^{(j)}$ with analyticity regions $\Omega_j$, which satisfy a
RHP on the contour shown on Figure 1:
\begin{equation}
\begin{split}
\chi^{(1)}(x,\lambda) &=\chi^{(2)}(x,\lambda) G(\lambda),
\qquad \lambda \in [-1,1] \\
\chi^{(4)}(x,\lambda) &=\chi^{(3)}(x,\lambda) G(\lambda),
\quad \lambda \in (-\infty,-1]\cup [1,\infty) \\
\chi^{(1)}(x,\lambda) &=\chi^{(3)}(x,\lambda) G(\lambda),
\qquad \lambda=e^{i \varphi},\quad \varphi\in(0,\pi)\\
\chi^{(4)}(x,\lambda) &=\chi^{(2)}(x,\lambda) G(\lambda),
\qquad \lambda=e^{i \varphi},\quad \varphi\in(\pi,2\pi)
\end{split}
\end{equation}
where the sewing function is given by
\[G(\lambda) =(S^-(\lambda))^{-1}S^+(\lambda).\]
The FAS obey the following symmetry relations:
\begin{equation}
\begin{aligned}
\left[\chi^{(1)}(x,\lambda^*)\right]^\dag &= [\chi^{(2)}(x,\lambda)]^{-1},
&\quad \left[\chi^{(4)}(x,\lambda^*)\right]^\dag &= [\chi^{(3)}(x,\lambda)]^{-1}\\
J_1\chi^{(1)}(x,-\lambda)J_1 &= \chi^{(2)}(x,\lambda),
&\quad J_1\chi^{(4)}(x,-\lambda)J_1 &= \chi^{(3)}(x,\lambda)\\
J_2\chi^{(1)}(x,1/\lambda)J_2 &= \chi^{(4)}(x,\lambda),
&\quad J_2\chi^{(2)}(x,1/\lambda)J_2 &= \chi^{(3)}(x,\lambda).
\end{aligned}
\end{equation}

\section{Recursion operators}

This section is dedicated to the construction of recursion
operator for the NLEE (\ref{nee}). For this to be done we
are applying the method proposed by G\"urses, Karasu and
Sokolov (GKS) \cite{sok}. The recursion operator is viewed
here a pseudo-differential operator to map a symmetry
of a NLEE into another symmetry. The underlying idea of the
GKS method is to interrelate two adjacent flows $\check{A}$
and $A$ with evolution variables $\tau$ and $t$ respectively
through a equality in the form \cite{sok, wang}
\begin{equation}
\check{V}=\left(\lambda^2 + \lambda^{-2}\right)V+B.
\label{checkv_v}\end{equation}
The matrix-valued functions $\tilde{V}$ and $V$
are involved in the Lax operators $\tilde{A}$ and $A$ as
follows
\begin{eqnarray}
\tilde{A}&=&i \partial_{\tau} + \tilde{V}(x,t,\lambda),\\
A&=&i \partial_t + V(x,t,\lambda)
\end{eqnarray}
All quantities above must be invariant under the action of
the $\bbbz_2$ reductions (\ref{eq:RC1})  and (\ref{red3}). This explains the choice of the coefficient before
$V$ in (\ref{checkv_v}) which is invariant under
$\lambda\to -\lambda$ and $\lambda\to \lambda^{-1}$.
The operator $B$ is chosen to be a rational function of
$\lambda$
\begin{equation}
B=B_0+\lambda B_1+\frac{1}{\lambda}B_{-1}+\lambda^2 B_{2}
+\frac{1}{\lambda^2}B_{-2},\qquad
B_{-k}=J_2B_{k}J_2,\quad k=1,2
\end{equation}
The matrices involved in $B$ are hermitian ones and have
a block structure given by
\begin{equation}
B_1\equiv \left(\begin{array}{cc}
0  & \mathbf{c}^T\\
\mathbf{c}^{*} & 0
\end{array}\right),\qquad
B_2\equiv \left(\begin{array}{cc}
d & 0 \\
0 & D
\end{array}\right),\qquad
D\equiv \left(\begin{array}{cc}
\alpha & \beta \\
\beta^* & \delta
\end{array}\right),
\label{LB_form}\end{equation}
in a complete analogy with the second Lax
operator (\ref{lax_2}).

At the begining of our consideration, let us remark
that the zero curvature condition of the Lax operators
$L$ and $A$ could be written in the following manner:
\begin{equation}
i L_t \equiv  [L,V],
\label{lax_orig}\end{equation}
where $L_t\equiv \lambda L_{1,t} + \lambda^{-1}L_{-1,t}$.
After substituting (\ref{checkv_v}) in the analog of (\ref{lax_orig}),
when the evolution paramenter $t$ is replaced by $\tau$, we
obtain the following basic equation:
\begin{equation}
i L_{\tau}= i \left(\lambda^2+\lambda^{-2}\right)
L_t + [L,B].
\label{recurop_rel}\end{equation}
Then the recursion operator $\mathcal{R}$ acts as follows:
\begin{equation}
\left(\begin{array}{c}
\mathbf{u}\\ \mathbf{u}^*
\end{array}\right)_{\tau}=\mathcal{R}\left(\begin{array}{c}
\mathbf{u} \\ \mathbf{u}^*
\end{array}\right)_t.
\label{recurs_op_def}\end{equation}
Since all matrices are traceless, we have
\[d=-\tr D=-(\alpha+\delta).\]
By comparing coefficients before different powers of $\lambda$,
one splits equation (\ref{recurop_rel}) into
\begin{eqnarray}
i L_{1,t} &+& [L_1,B_2] =0 \qquad\Rightarrow\qquad
i \mathbf{u}_t+(D^*-d)\mathbf{u} = 0,\label{recurel_1}\\
i B_{2,x} &+& [L_1,B_1] =0 \quad\Rightarrow\quad
\begin{array}{cc}
i D_{x} + \mathbf{u}^{*}\mathbf{c}^{T}
- \mathbf{c}^{*}\mathbf{u}^{T} = 0\\
i d_{x} + \mathbf{u}^{T}\mathbf{c}^{*}
- \mathbf{c}^{T}\mathbf{u}^{*} = 0
\end{array} \label{recurel_2}\\
i L_{1,\tau} &=& i L_{-1,t} + i B_{1,x}
+ [L_1,B_0] + [L_{-1},B_2],\label{recurel_3a}\\
i B_{0,x} &+& [L_1,B_{-1}]+[L_{-1},B_{1}] = 0.
\label{recurel_4a}
\end{eqnarray}
The rest of relations could be obtained from those above, by
multiplying each quantity by $J_2$ on its left and on
its right hand side.

The system (\ref{recurel_1}) is linear for the matrix elements
of $D$. A solution to (\ref{recurel_1}) is given by
\begin{equation}
\begin{split}
\alpha &= i (uu^*_t+v^*v_t) - d(2|v|^2-|u|^2)\qquad \beta=i (vu^*_t-u^*v_t) +3d u^*v\\
\delta &=-i (uu^*_t+v^*v_t)- d (2|u|^2-|v|^2).
\label{greek}
\end{split}\end{equation}
The solvability condition for the system of linear equations (\ref{greek}) is:
\begin{equation}
\alpha_x |v|^2 + \delta_x |u|^2 - \beta_x uv^* - \beta^*_xu^*v = 0
\label{constr_J}\end{equation}
One can check that
\begin{equation}
\mathbf{c}=\left(\begin{array}{c}
c\\s\end{array}\right)=
\frac{i }{2}\left(\begin{array}{c}
u(\delta_x-\alpha_x)-2v\beta^{*}_x\\
-v(\delta_x-\alpha_x)-2u\beta_x\end{array}\right) + \kappa \left(\begin{array}{c} u\\ v\end{array}\right)
\label{c_result}\end{equation}
where $\kappa$ is real, but otherwise arbitrary, is the general solution of (\ref{greek}). In what follows we shall set $\kappa =0$ without any loss of
generality.  It could be written in the following matrix form
\begin{equation}
\left(\begin{array}{c}
c\\ s\\ c^* \\ s^*
\end{array}\right)_{x}=\mathcal{A}
\left(\begin{array}{c}
\beta \\ \beta^* \\
\frac{\alpha-\delta}{2}\\
\frac{\alpha+\delta}{2}
\end{array}\right),\qquad
\mathcal{A}:=i \frac{\rmd}{d x}
\left(\begin{array}{cccc}
0   & -v  & -u   & 0 \\
-u  & 0   & v    & 0 \\
v^* & 0   & u^*  & 0 \\
0   & u^* & -v^* & 0
\end{array}\right)\frac{\rmd}{d x}.
\label{recurop_factor1}\end{equation}
Next step is to find the function $d$, involved in the
expressions for $\alpha$, $\beta$ etc. For this to be
done, we make use of the condition (\ref{constr_J}) which
leads to a linear differential equation for $d$:
\begin{eqnarray}
-2d_x &+&   \rmi(u^*u_t+v^*v_t)_x-i \left[(uu^*_t+v^*v_t)(|v|^2-|u|^2)_x
\right.\nonumber\\
&-&\left.(uv^*_t-v^*u_t)(u^*v)_x+(u^*v_t-v^*u_t)(uv^*)_x\right]=0.
\end{eqnarray}
After some simple manipulations one obtains
\begin{equation}
d= -\frac{i }{2}(uu^*_t+vv^*_t) -
\frac{i }{2}\partial^{-1}_x\left(u_tu^*_x+v_tv^*_x
-u^*_tu_x-v^*_tv_x\right),
\label{J_expr}\end{equation}
where $\partial^{-1}_x:=\int^{x}_{\pm\infty}d y$.

To calculate $\mathcal{R}$, one needs the diagonal matrix
$B_0$. It directly follows from (\ref{recurel_4a}) that
the following equation holds true
\begin{equation}
i  B_{0,x}  = 2\left(\begin{array}{ccc}
uc^*-u^*c+v^*s-vs^* & 0 & 0 \\
0 & u^*c-uc^* & 0 \\
0 &     0     & vs^*-v^*s
\end{array}\right).
\end{equation}
Taking into account formulas (\ref{recurel_2}), one deduces that
\[B_0= 2 \left(\begin{array}{ccc}
\alpha-\delta & 0   & 0   \\
0   & -\alpha & 0   \\
0   & 0   & \delta
\end{array}\right).\]
Furthermore, taking into account the structure of (\ref{recurel_3a}),
one can easily see that $\mathcal{R}$ can be split in the following form:
\begin{equation}
\mathcal{R}= \mathcal{C} + \mathcal{R}_0 +\mathcal{D}.
\end{equation}
The first term  $\mathcal{C}$ originates from the first term in
(\ref{recurel_3a}), and it is a constant matrix $\diag(-1, 1, -1,
1)$. The third operator $\mathcal{D}$ is obtained from the two
commutators in (\ref{recurel_3a}). It turns out that it splits
into a local term
\[\mathcal{D}_{\rm{loc}} :=
\left(\begin{array}{cccc}
8|u|^2-1 & -4u^2(3|v|^2-1) & 0 & 12uv|u|^2\\
0 & -12uv|v|^2 & 1-8|v|^2 & 4v^2(3|u|^2-1)\\
-4(u^*)^2(3|v|^2-1) & 8|u|^2-1 & 12u^*v^*|u|^2 & 0 \\
-12u^*v^*|v|^2 & 0 & 4(v^*)^2(3|u|^2-1) & 1-8|v|^2
\end{array}\right)\]
and a nonlocal one
\[\mathcal{D}_{\rm{nonl}} =
4\left(\begin{array}{c}
u\left(1-3|v|^2\right) \\ v\left(3|u|^2-1\right) \\
u^*\left(3|v|^2-1\right) \\ v^*\left(1-3|u|^2\right)
\end{array}\right)
\partial^{-1}_x\left[(u^*_x,v^*_x,-u_x,-v_x) . \right].\]
Finally the operator $\mathcal{R}_0=\mathcal{A}
(\mathcal{B}_{\rm{loc}}+\mathcal{B}_{\rm{nonl}})$
splits into a local and a nonlocal parts as follows:
\begin{eqnarray}
\mathcal{B}_{\rm loc} &=& \frac{i }{4} \left(\begin{array}{cccc}
0  & -4u^* & 2v(3|v|^2-1) & -6u^*v^2 \\
2v^*(1-3|v|^2) & 6u(v^*)^2 & 0 &  4u  \\
-4u^* & 0 & 3u(|v|^2-|u|^2) &  -v(1+6|u|^2) \\
- u^* & - v^* & 0 &  0 \end{array}\right),\label{local_part}\\
\mathcal{B}_{\rm nonl} &=& - \frac{3i }{4}
\left(\begin{array}{c}
2u^*v \\ 2uv^* \\
|u|^2-|v|^2 \\
-1/3
\end{array}\right)
\partial^{-1}_x\left[(u^*_x,v^*_x,-u_x,-v_x) \right].
\label{nonlocal_part}\end{eqnarray} In fact, $\mathcal{R}_0$
represents the recursion operator in the case of polynomial bundle
Lax pair \cite{JGSP}:
\begin{eqnarray*}
L&:=&i \partial_x + \lambda L_1,\\
A&:=&i \partial_t + \lambda A_1 +
\lambda^2 A_2.
\end{eqnarray*}

\section{Recursion operators and `squared solutions'}

This approach is based on the Wronskian relations, mapping the potential of the Lax operator $L$ onto the scattering data, which allow one to
introduce the `squared solutions' for $L$ (see, e.g. \cite{GVY*08} and the references therein).

The Wronskian relations, derived in \cite{ours}, take the form:
\begin{equation}\label{eq:Wr2'}
\begin{aligned}
\left. \left\langle i \chi^{-1} J_0 \chi (x,\lambda)- i J_0, E_\alpha\right\rangle \right|_{x=-\infty}^{\infty}
 &=  \int_{x=-\infty}^{\infty} dx\, \left\langle [L_1,J_0] , \Phi_1(x,\lambda) \right\rangle \\
 \left. i \left\langle \chi^{-1} \delta \chi (x,\lambda),E_\alpha \right\rangle
\right|_{x=-\infty}^{\infty} &=
 -\int_{x=-\infty}^{\infty} dx\, \left\langle  \delta L_1, \Phi_1 (x,\lambda) \right\rangle,
\end{aligned}
\end{equation}
where we have  introduced the squared solutions:
\begin{equation}\label{eq:ssol4}
\Phi_1(x,\lambda) = \lambda e_\alpha(x,\lambda) +  \lambda^{-1} \varphi_0(e_\alpha)(x,\lambda),
\end{equation}
and
\begin{equation}\label{eq:R2.0}
e_\alpha(x,\lambda) = \chi E_\alpha \chi^{\bf -1}(x,\lambda).
\end{equation}
where $E_\alpha$ is one of the Cartan-Weyl generators.

In this approach, see \cite{AKNS*74,GVY*08}, one  picks up a
certain `projections' of the `squared solutions' (see $
K_{1;\alpha}^{\pm,\perp} (x,\lambda)$ in eq. (\ref{eq:R4.6})
below) which contribute to the right hand sides of eq.
(\ref{eq:Wr2'}). Then the recursion operators can be introduced as
the ones, that have $K_{1;\alpha}^{\pm,\perp} (x,\lambda)$ as
eigenfunctions.

Let us introduce
\begin{equation}\label{eq:R1.1}
J_1 =\diag(1,-1,-1), \qquad K_1 =\diag(1,0,-1), \qquad K_0 = \diag(1,-1,1)
\end{equation}
and
\begin{equation}\label{eq:Phiak}
\Phi_{k;\alpha}^\pm (x,\lambda) = \lambda^k e_\alpha^\pm (x,\lambda) + \lambda^{-k} \varphi_0(e_\alpha^\pm (x,\lambda)),
\end{equation}
which satisfy the equations:
\begin{equation}\label{eq:R2.2}
\begin{aligned}
i \frac{\partial \Phi_{k;\alpha} }{\partial x} &+ [L_1, \Phi_{k+1;\alpha}^\pm (x,\lambda)] +
 [L_{-1}, \Phi_{k-1;\alpha}^\pm (x,\lambda)] =0\\
\varphi_0 (\Phi_{k;\alpha}^\pm (x,\lambda) )&= \Phi_{-k;\alpha}^\pm (x,\lambda) , \qquad
\varphi_0 (\Phi_{0;\alpha}^\pm (x,\lambda) )= \Phi_{0;\alpha}^\pm (x,\lambda) \\
i \frac{\partial \Phi_{0;\alpha} }{\partial x} & + (\openone + \varphi_0)[L_1, \Phi_{1;\alpha}^\pm (x,\lambda)] =0
\end{aligned}
\end{equation}
In addition, we have:
\begin{equation}\label{eq:R2.4}
\begin{aligned}
(\lambda +\lambda^{-1}) \Phi_{k;\alpha}^\pm (x,\lambda) &= \Phi_{k+1;\alpha}^\pm (x,\lambda) + \Phi_{k-1;\alpha}^\pm (x,\lambda)
\end{aligned}
\end{equation}
In what follows we will use only the equations for $\Phi_{0;\alpha}^\pm (x,\lambda)$ and $\Phi_{1;\alpha}^\pm (x,\lambda)$:
\begin{equation}\label{eq:R2.7}
\begin{aligned}
i \frac{\partial \Phi_{0;\alpha} }{\partial x}  + (\openone + \varphi_0)[L_1, \Phi_{1;\alpha}^\pm (x,\lambda)] &=0 \\
i \frac{\partial \Phi_{1;\alpha} }{\partial x}  -  [L_1 - L_{-1}, \Phi_{0;\alpha}^\pm (x,\lambda)] &= -(\lambda +\lambda^{-1})
 [L_1 , \Phi_{1;\alpha}^\pm (x,\lambda)]
\end{aligned}
\end{equation}
Next, we insert the splitting:
\begin{equation}\label{eq:R3.1}
\begin{aligned}
 \Phi_{0;\alpha}^\pm (x,\lambda) &=  H_{0;\alpha}^\pm (x,\lambda) + K_{0;\alpha}^\pm (x,\lambda) , &\quad
\Phi_{1;\alpha}^\pm (x,\lambda) &=  H_{1;\alpha}^\pm (x,\lambda) + K_{1;\alpha}^\pm (x,\lambda) \\
H_{0;\alpha}^\pm (x,\lambda) &= \varphi_0(H_{0;\alpha}^\pm (x,\lambda)) , &\quad
K_{0;\alpha}^\pm (x,\lambda) &= \varphi_0(K_{0;\alpha}^\pm (x,\lambda))
\end{aligned}
\end{equation}
and obtain
\begin{equation}\label{eq:R4.1}
\begin{aligned}
i \frac{\partial H_{0;\alpha}^\pm }{\partial x} &+ (\openone + \varphi_0) [L_1, K_{1;\alpha}^\pm (x,\lambda)] =0 \\
i \frac{\partial K_{0;\alpha}^\pm }{\partial x} &+ (\openone + \varphi_0) [L_1, H_{1;\alpha}^\pm (x,\lambda)] =0 \\
i \frac{\partial H_{1;\alpha}^\pm }{\partial x} &- [L_1 -L_{-1} , K_{0;\alpha}^\pm (x,\lambda)] =-(\lambda +\lambda^{-1})
[L_1 ,K_{1;\alpha}^\pm (x,\lambda)] \\
i \frac{\partial K_{1;\alpha}^\pm }{\partial x} &- [L_1 -L_{-1} , H_{0;\alpha}^\pm (x,\lambda)] =-(\lambda +\lambda^{-1})
[L_1, H_{1;\alpha}^\pm (x,\lambda)]
\end{aligned}
\end{equation}
The integration the first two of the above equations gives:
\begin{equation}\label{eq:R4.5}
\begin{aligned}
H_{0;\alpha}^\pm (x,\lambda) & = h_{00;\alpha}^\pm + i (\openone + \varphi_0) \partial_{x}^{-1} [L_1 , K_{1;\alpha}^{\pm,\perp} (x,\lambda)] \\
K_{0;\alpha}^\pm (x,\lambda) & = k_{00;\alpha}^\pm + i (\openone + \varphi_0) \partial_{x}^{-1} [L_1 , H_{1;\alpha}^{\pm,\perp} (x,\lambda)],
\end{aligned}
\end{equation}
where $h_{00;\alpha}^\pm$ and $k_{00;\alpha}^\pm$ are matrix-valued constants.
  Thus we expressed $K_{0;\alpha}^{\pm } $ and $H_{0;\alpha}^{\pm } $ in
terms of $K_{1;\alpha}^{\pm ,\perp} $ and $H_{1;\alpha}^{\pm ,\perp} $ where
\begin{equation}\label{eq:R4.6}
\begin{aligned}
H_{1;\alpha}^{\pm, \perp} (x,\lambda) &= H_{1;\alpha}^{\pm} (x,\lambda) - \frac{3}{2} L_2 \langle H_{1;\alpha}^\pm ,L_2\rangle  \\
K_{1;\alpha}^{\pm , \perp} (x,\lambda) &= K_{1;\alpha}^{\pm} (x,\lambda) - \frac{1}{2} L_1 \langle K_{1;\alpha}^\pm ,L_1\rangle
\end{aligned}
\end{equation}
 where $L_2 =L_1^2 - 2/3 \openone$. Let us now derive the
equations for  $K_{1;\alpha}^{\pm ,\perp} $ and $H_{1;\alpha}^{\pm
,\perp} $. To do this, we insert eq. (\ref{eq:R4.6} and insert it
into the last two equations of (\ref{eq:R4.1}). Thus we get:
\begin{equation}\label{eq:R4.7}
\begin{aligned}
i \frac{\partial H_{1;\alpha}^{\pm,\perp} }{\partial x} &+ \frac{3i}{2} L_{2,x} \langle H_{1;\alpha}^\pm ,L_2\rangle
+\frac{3 i}{2} L_2 \frac{\partial   \langle H_{1;\alpha}^\pm ,L_2\rangle }{\partial x} - [L_1 -L_{-1} , K_{0;\alpha}^\pm (x,\lambda)]
 \\ &=-(\lambda +\lambda^{-1}) [L_1 ,K_{1;\alpha}^\pm (x,\lambda)] \\
i \frac{\partial K_{1;\alpha}^{\pm,\perp} }{\partial x} &+ \frac{i}{2} L_{1,x} \langle K_{1;\alpha}^\pm ,L_1\rangle
+\frac{i}{2} L_1 \frac{\partial   \langle K_{1;\alpha}^\pm ,L_1\rangle}{\partial x}  - [L_1 -L_{-1} , H_{0;\alpha}^\pm (x,\lambda)]
 \\ &=-(\lambda +\lambda^{-1}) [L_1 ,H_{1;\alpha}^\pm (x,\lambda)].
\end{aligned}
\end{equation}
Next, we separate the left hand sides of eq. (\ref{eq:R4.7}) into terms commuting with $L_1$ and $L_2$ and terms `orthogonal' to them.
In order to calculate the coefficients $\langle H_{1;\alpha}^\pm ,L_2\rangle$ and $\langle K_{1;\alpha}^\pm ,L_1\rangle$, we
multiply both sides of the first (resp. the second) of the eqs. (\ref{eq:R4.7}) by $L_2$ (resp. by $L_1$) and take the trace.
The result is:
\begin{equation}\label{eq:R5.0}
\begin{aligned}
i \left\langle \frac{\partial H_{1;\alpha}^{\pm,\perp} }{\partial x}, L_2\right\rangle &
+i \frac{\partial   \langle H_{1;\alpha}^\pm ,L_2\rangle}{\partial x}  -  \langle [L_1 -L_{-1} , K_{0;\alpha}^\pm (x,\lambda)], L_2\rangle =0
  \\
i \left\langle \frac{\partial K_{1;\alpha}^{\pm,\perp} }{\partial x} , L_1 \right\rangle &+
i \frac{\partial   \langle K_{1;\alpha}^\pm ,L_1\rangle}{\partial x}  - \langle [L_1 -L_{-1} , H_{0;\alpha}^\pm (x,\lambda)], L_1 \rangle =0
\end{aligned}
\end{equation}
Integrating (\ref{eq:R5.0}) we get:
\begin{equation}\label{eq:01}
\begin{aligned}
 \langle H_{1;\alpha}^\pm ,L_2\rangle &= h_{01,\alpha}^\pm -\partial_{x}^{-1} \left\langle \frac{\partial H_{1;\alpha}^{\pm,\perp} }{\partial x}, L_2\right\rangle -i \partial_{x}^{-1} \langle [L_1 -L_{-1} , K_{0;\alpha}^\pm (x,\lambda)], L_2\rangle \\
 \langle K_{1;\alpha}^\pm ,L_1\rangle &= k_{01,\alpha}^\pm -\partial_{x}^{-1} \left\langle \frac{\partial K_{1;\alpha}^{\pm,\perp} }{\partial x}, L_1\right\rangle ,
\end{aligned}
\end{equation}
where $k_{01}^\pm $ and $h_{01}^\pm $ are matrix-valued integration constants.

Skipping the details we obtain:
\begin{equation}\label{eq:R22.1}
\begin{aligned}
\Lambda_1 K_{1;\alpha}^{\pm,\perp} &= (\lambda + \lambda^{-1}) H_{1;\alpha}^{\pm,\perp}
- \ad_{L_1}^{-1} \left( [L_1 -L_{-1} ,h_{00}] + \frac{3}{2}L_{2,x} \partial_x^{-1} \langle [L_{2},L_{-1}] ,k_{00} \rangle \right) \\
\Lambda_2 H_{1;\alpha}^{\pm,\perp} &= (\lambda + \lambda^{-1}) K_{1;\alpha}^{\pm,\perp} + \frac{3i}{2} \ad_{L_1}^{-1}
\left( L_{2,x} h_{01} + i L_{2,x} \partial_x^{-1} \langle [L_2,L_{-1}],k_{00}\rangle \right) \\
&- \ad_{L_1}^{-1} \left( [L_1 -L_{-1} ,k_{00}] \right)
\end{aligned}
\end{equation}
where $h_{00}$ and $k_{00}$ are arbitrary constants which we shall
set to be equal to zero. The operators  $\Lambda_j$, $j=1,2$ are defined as follows:
\begin{equation}\label{eq:R22.2}
\begin{aligned}
\Lambda_1 K_{1;\alpha}^{\pm,\perp} &= -i \ad_{L_1}^{-1} \left( \frac{\partial K_{1;\alpha}^{\pm,\perp} }{\partial x}
- (\openone -  \varphi_0) \left[ L_1 -L_{-1} , \partial_x^{-1} [L_1, K_{1;\alpha}^{\pm,\perp}] \right] \right) \\
&+ \frac{i}{2} \ad_{L_1}^{-1} L_{1,x} \partial_x^{-1} \left\langle  \frac{\partial K_{1;\alpha}^{\pm,\perp} }{\partial x} , L_1 \right\rangle \\
\Lambda_2 H_{1;\alpha}^{\pm,\perp} &= -i \ad_{L_1}^{-1} \left( \frac{\partial H_{1;\alpha}^{\pm,\perp} }{\partial x}
-  (\openone - \varphi_0) \left[ L_1 -L_{-1} , \partial_x^{-1} [L_1, H_{1;\alpha}^{\pm,\perp}] \right] \right.  \\
& \left. -\frac{3}{2}   L_{2,x}
\partial_x^{-1} \left( \left\langle  \frac{\partial H_{1;\alpha}^{\pm,\perp} }{\partial x} , L_2 \right\rangle
+ \left\langle [L_2 - L_{-2},L_1 -L_{-1} ], \partial_x^{-1} [L_1, H_{1;\alpha}^{\pm,\perp}] \right\rangle \right) \right)
\end{aligned}
\end{equation}
Thus,  we obtain:
\begin{equation}\label{eq:R22.3}
\begin{aligned}
\Lambda_2\Lambda_1 K_{1;\alpha}^{\pm,\perp} &= (\lambda +\lambda^{-1})^2 K_{1;\alpha}^{\pm,\perp} \\
\Lambda_1\Lambda_2 H_{1;\alpha}^{\pm,\perp} &= (\lambda +\lambda^{-1})^2 H_{1;\alpha}^{\pm,\perp}
\end{aligned}
\end{equation}

\section{Conclusions}

We derived the recursion operators $\Lambda$ for the system  (\ref{nee}) related to
the symmetric space $SU(3)/S(U(1)\times U(2))$ with an additional $\bbbz_2$-reduction.
Our first  derivation of $\Lambda$ is based on the GKS method \cite{sok}.

Another way to construct $\Lambda$  consists in using the Wronskian relations.
They allow us to determine the `squared solutions' of $L$ which are treated as
eigenfunctions of $\Lambda$.

Our results could be extended for operators $L$ related to
a generic symmetric space of the type
$\mathbf{A.III}\cong SU(n+k)/S(U(n)\times U(k))$ as well as to
 other types of symmetric spaces.

\section*{Acknowledgements}

The authors acknowledge support from the Royal Society and
the Bulgarian academy of sciences via joint research project
"Reductions of Nonlinear Evolution Equations and analytic
spectral theory". V.S.G. and G.G.G. are thankful to the
organizers of the  international workshop: ``Nonlinear
Physics: Theory and Experiment. VI'' for the financial support and the warm
hospitality in Gallipoli. The work of G.G.G. is supported by the
Science Foundation of Ireland (SFI), under Grant No.
09/RFP/MTH2144.

\appendix
\section{Some intermediate results}

In this appendix we will present some intermediate results, used in section 4. The involutions $J$ induce a $\bbbz_2$ grading in the Lie algebra ${\frak g}$ as follows:
\begin{equation}\label{eq:R1.2}
\begin{aligned}
\mathfrak{g} &\equiv \mathfrak{g}^{(0)} \oplus  \mathfrak{g}^{(1)} , \\
\mathfrak{g}^{(0)} &\equiv \{ Y\colon Y=JYJ^{-1} \}, &\qquad
\mathfrak{g}^{(1)} &\equiv \{ X\colon X= -JXJ^{-1} \} \\
Y &= \left(\begin{array}{ccc} -k-n & 0 & 0 \\ 0 & k & m \\ 0 & m^* & n \end{array}\right) , &\qquad
X &= \left(\begin{array}{ccc} 0 & a & b \\ a^* & 0 & 0 \\ b^* & 0 & 0 \end{array}\right)
\end{aligned}
\end{equation}
The inner automorphism $\varphi_0$ staying in the formula for the ``extended squared solutions''  (\ref{eq:R3.1}) can be represented in the form:
\begin{equation}\label{eq:R1.3}
\begin{aligned}
\varphi_0(Z) &\equiv K_0 Z K_0^{-1} ,   \\
Y &= Y^+ + Y^-, & Y^+ &= \left(\begin{array}{ccc} -k-n & 0 & 0 \\ 0 & k & 0 \\ 0 & 0 & n \end{array}\right) ,
&  Y^- &= \left(\begin{array}{ccc} 0 & 0 & 0 \\ 0 & 0 & m \\ 0 & m^* & 0 \end{array}\right) ,\\
X &= X^+ + X^-, &  X^+ &= \left(\begin{array}{ccc} 0 & 0 & b \\ 0 & 0 & 0 \\ b^* & 0 & 0 \end{array}\right) ,
&  X^- &= \left(\begin{array}{ccc} 0 & a & 0 \\ a^* & 0 & 0 \\ 0 & 0 & 0 \end{array}\right) ,
\end{aligned}
\end{equation}
Next we take into account that $\varphi_0(H_0)=H_0$ and  $\varphi_0(K_0)=K_0$ and therefore:
\begin{equation}\label{eq:R5.1}
\begin{aligned}
\langle [L_1 -L_{-1} , K_{0;\alpha}^\pm (x,\lambda)], L_2\rangle &= 2|u|^2 (\kappa_{0,\alpha}^\pm v^* - \kappa_{0,\alpha}^{\pm,*} v ) \\
\langle [L_1 -L_{-1} , H_{0;\alpha}^\pm (x,\lambda)], L_1\rangle &=0
\end{aligned}
\end{equation}
where
\begin{equation}\label{eq:KH0}
K_{0,\alpha}^\pm  = \left(\begin{array}{ccc} 0 & 0 & \kappa_{0,\alpha}^\pm \\ 0 & 0 & 0 \\
\kappa_{0,\alpha}^{\pm,*} & 0 & 0 \end{array}\right) , \qquad
H_{0,\alpha}^\pm  = \left(\begin{array}{ccc} -k_{0,\alpha}^\pm -n_{0,\alpha}^\pm & 0 & 0 \\ 0 & k_{0,\alpha}^\pm & 0 \\
0 & 0 & n_{0,\alpha}^\pm \end{array}\right) ,
\end{equation}
Introduce the notations
\begin{equation}\label{eq:R5.a}
\begin{aligned}
K_{1,\alpha}^{\pm} = \left(\begin{array}{ccc} 0 & w_{1,\alpha}^{\pm} & z_{1,\alpha}^{\pm} \\
w_{1,\alpha}^{\pm,*} & 0 & 0 \\ z_{1,\alpha}^{\pm,*} & 0 & 0 \end{array}\right) , \qquad
H_{1,\alpha}^{\pm} = \left(\begin{array}{ccc} - k_{1,\alpha}^{\pm} - n_{1,\alpha}^{\pm} & 0 & 0 \\
0 & k_{1,\alpha}^{\pm} & m_{1,\alpha}^{\pm} \\ 0 & m_{1,\alpha}^{\pm,*} & n_{1,\alpha}^{\pm} \end{array}\right)
\end{aligned}
\end{equation}
Then
\begin{equation}\label{eq:R5.6}
\begin{aligned}
 \langle H_{1;\alpha}^\pm ,L_2\rangle &= h_{01,\alpha}^\pm -\partial_{x}^{-1} \left\langle \frac{\partial H_{1;\alpha}^{\pm,\perp} }{\partial x}, L_2\right\rangle -2i \partial_{x}^{-1} |u|^2 ( v^* k_{00,\alpha}^\pm -  v k_{00,\alpha}^{\pm,*} ) \\
&+ 8 \partial_{x}^{-1} |u|^2 \left( v^*\partial_{x}^{-1} (u m_{1,\alpha}^\pm) + v\partial_{x}^{-1}( u^* m_{1,\alpha}^{\pm,*})\right)\\
 \langle K_{1;\alpha}^\pm ,L_1\rangle &= k_{01,\alpha}^\pm -\partial_{x}^{-1} \left\langle \frac{\partial K_{1;\alpha}^{\pm,\perp} }{\partial x}, L_1\right\rangle
\end{aligned}
\end{equation}

\end{document}